\documentclass[usenatbib]{mn2e}
\usepackage{graphicx,times}
\usepackage{bm}
\graphicspath{{./fig/}{./png/}}
\newcommand{\mean}[1]{\overline{#1}}
\newcommand{\meanv}[1]{\overline{\bm #1}}
\newcommand{\pd}{\partial}
\newcommand{\cst}{c_{\rm s}}
\newcommand{\urms}{u_{\rm rms}}

\newcommand{\brms}{B_{\rm rms}}

\newcommand{\Beq}{B_{\rm eq}}

\newcommand{\Pm}{{\rm Pm}}
\newcommand{\Rey}{{\rm Re}}
\newcommand{\Rem}{{\rm Rm}}
\newcommand{\Cem}{{\rm Cm}}

\newcommand{\aSS}{{\alpha_{\rm SS}}}
\newcommand{\nut}{{\nu_{\rm t}}}
\newcommand{\rxy}{R_{xy}}
\newcommand{\mxy}{M_{xy}}

\def\onehalf{{\textstyle{1\over2}}}
\def\onethird{{\textstyle{1\over3}}}
\newcommand{\Fig}[1]{Fig.~\ref{#1}}

\newcommand{\Eq}[1]{equation~(\ref{#1})}

\newcommand{\yaj}[3]{ #1, {AJ,} {#2}, #3}

\newcommand{\yapj}[3]{ #1, {ApJ,} {#2}, #3}

\newcommand{\yan}[3]{ #1, {Astron.\ Nachr.,} {#2}, #3}

\newcommand{\yana}[3]{ #1, {A\&A,} {#2}, #3}

\newcommand{\ypepi}[3]{ #1, {Phys.\ Earth Planet.\ Int.,} {#2}, #3}

\newcommand{\yjetp}[3]{ #1, {Sov.\ Phys.\ JETP,} {#2}, #3}

\newcommand{\yprl}[3]{ #1, {Phys.\ Rev.\ Lett.,} {#2}, #3}

\newcommand{\ymn}[3]{ #1, {MNRAS,} {#2}, #3}
\newcommand{\ynjp}[3]{ #1, {NJP,} {#2}, #3}

\newcommand{\ypre}[3]{ #1, {Phys.\ Rev.\ E,} {#2}, #3}
\newcommand{\ycpc}[3]{ #1, {Comp.\ Phys.\ Comm.,} {#2}, #3}
\newcommand{\ypnas}[3]{ #1, {Proc.\ Nat.\ Acad.\ Sci.,} {#2}, #3}
\newcommand{\ypaipc}[3]{ #1, {AIPC,} {#2}, #3}

\newcommand{\yjour}[4]{ #1, {#2}, {#3}, #4}

\newcommand{\ybook}[3]{ #1, {#2} (#3)}

\title[MRI-driven dynamos at low Pm]
{Magnetorotational instability driven dynamos at low magnetic Prandtl numbers}
\author[P.\ J.\ K\"apyl\"a \& M.\ J.\ Korpi]{P.\ J.\ K\"apyl\"a$^{1,2}$ 
and M.\ J.\ Korpi$^{1}$\\
$^1$Department of Physics, Gustaf H\"allstr\"omin katu 2a (PO Box 64),
FI-00064 University of Helsinki, Finland\\
$^2$NORDITA, AlbaNova University Center, Roslagstullsbacken 23,
SE-10691 Stockholm, Sweden
}
\date{Accepted 2010 December 9. Received 2010 November 9; in original form 2010 April 14}

\begin{document}
\maketitle

\begin{abstract}
  Numerical simulations of the magnetorotational instability (MRI)
  with zero initial net flux in a non-stratified isothermal cubic
  domain are used to demonstrate the importance of magnetic boundary
  conditions.In fully periodic systems the level of turbulence
  generated by the MRI strongly decreases as the magnetic Prandtl
  number ($\Pm$), which is the ratio of kinematic viscosity and
  magnetic diffusion, is decreased. No MRI or dynamo action below
  $\Pm=1$ is found, agreeing with earlier investigations.  Using
  vertical field conditions, which allow the generation of a net
  toroidal flux and magnetic helicity fluxes out of the system, the
  MRI is found to be excited in the range $0.1\le\Pm\le10$, and that
  the saturation level is independent of $\Pm$. In the vertical field
  runs strong mean-field dynamo develops and helps to sustain the MRI.
\end{abstract}

\label{firstpage}

\section{Introduction}
The realization of the astrophysical signifigance of the
magnetorotational instability \citep{BH91}, first discovered in the
context of Couette flow \citep{V59,C60}, seemed to resolve the
long-standing problem of the mechanism driving turbulence in accretion
disks. Early numerical simulations produced sustained turbulence,
large-scale magnetic fields and outward angular momentum transport
\citep[e.g.][]{BNST95,HGB95}.  These results also showed that a
significant qualitative difference exists between models where an
imposed uniform magnetic field is present as opposed to the situations
where such field is absent: the saturation level of turbulence and
angular momentum transport are substantially higher when a non-zero
vertical net flux is present \citep[e.g.][]{BNST95,Sea96}.  Also the
presence of an imposed net toroidal field seemed to enhance the
transport \citep{Sea96}.

In the meantime, a lot of numerical work has been done with zero net
flux setups that omit stratification and adopt fully periodic or
perfectly conducting boundaries in order to study the saturation
behaviour of the MRI in the simplest possible setting
\citep[e.g.][]{FP07,FPLH07,LKKBL09,KKV10}. Due to the boundary
conditions, the initial net flux in conserved and no magnetic helicity
fluxes out of the system are allowed.  The results of these
investigations have shown that as the numerical resolution of the
simulations increases, or equivalently as the explicit diffusion
decreases, the level of turbulence and angular momentum transport
transport decrease, constituting a convergence problem for zero net
flux MRI \citep{FPLH07}.  Runs with explicit diffusion show that
sustaining turbulence becomes increasingly difficult as the magnetic
Prandtl number, $\Pm=\nu/\eta$, where $\nu$ is the viscosity and
$\eta$ the magnetic diffusivity, is decreased
\citep{FPLH07}. Currently the convergence problem is without a
definite solution. It has been suggested that this issue could be
related to the $\Pm$-dependence of the fluctuation dynamo
\citep[e.g.][]{Schekea07}.  It has even been argued that the MRI in
periodic zero net flux systems would vanish in the limit of large
Reynolds numbers and that a large-scale dynamo would be needed to
sustain the MRI and turbulence \citep{V09}.  Notably, large-scale
dynamos have no problems operating at low magnetic Prandtl numbers as
long as the relevant Reynolds and dynamo numbers exceed critical
values \citep{B09}.

From the point of view of mean-field dynamo theory \citep{BS05},
systems with fully periodic or perfectly conducting boundaries are
rather special. In such closed setups magnetic helicity, defined as a
volume integral of $\bm{A}\cdot\bm{B}$, where $\bm{A}$ is the vector
potential and $\bm{B}=\bm\nabla\times\bm{A}$ is the magnetic field, is
a conserved quantity in ideal MHD. In the presence of magnetic
diffusion, magnetic helicity can change only on a timescale based on
microscopic diffusivity, which is usually a very long in any
astrophysical setting. Such a behaviour, which has been captured in
numerical simulations \citep{B01}, is well described by simple
mean-field models taking into account magnetic helicity conservation
\citep[e.g.][]{BB02}. This would mean that generating appreciable
large-scale magnetic fields, which are possibly vital for sustaining
the MRI, can take a very long time. Furthermore, the saturation value
of the mean magnetic field decreases inversely proportional to the
magnetic Reynolds number \citep[e.g.][]{CH96,B01}. In dynamo theory
this detrimental effect to the large-scale dynamo is known as the
catastrophic quenching \citep{VC92}.

The situation, however, changes dramatically if magnetic helicity flux
out of the system is allowed. In particular, the \cite{VC01} flux,
which requires large-scale velocity shear to be present and flows
along the isocontours of shear, is a potential mechanism that can
drive a magnetic helicity flux out of the system and alleviate
catastrophic quenching.  Indirect evidence for its importance exists
from convection simulations in a shearing box setup
\citep{KKB08,KKB10b}, where dynamo excitation is easier in systems
with boundaries that allow a net magnetic helicity flux. However,
these results can be explained by a somewhat higher critical dynamo
number in the perfect conductor case \citep{KKB10b}, which is a purely
kinematic effect. More dramatic differences between different boundary
conditions are seen in the nonlinear saturation regime, with strong
quenching of large-scale magnetic fields in the perfect conductor case
\citep{KKB10b}. The reason for this behaviour is not yet clear,
especially in light of recent results of \cite{HB10} who failed to
find evidence of the Vishniac--Cho flux in a numerical setup similar
to ours.

In the present paper we demonstrate that the boundary conditions play
a crucial role for the excitation of the MRI and the associated
large-scale dynamo. Following previous work that has shown that open
boundary conditions allow more efficient dynamo action
\citep{KKB08,KKB10b}, we model a system that is isothermal,
non-stratified, and the magnetic field has a zero net flux
initially. We then apply vertical field boundary conditions which
allow a magnetic helicity flux through the vertical boundaries by
letting the magnetic field cross them. We show that if the MRI is
excited, a large-scale dynamo is also excited and that the saturation
level of the turbulence, large-scale magnetic field, and angular
momentum transport are essentially independent of $\Pm$.  This is
contrasted by periodic simulations where we find a strong
$\Pm$-dependence in accordance with earlier studies. Our results also
suggest that for a given $\Pm$ the results (level of turbulence and
angular momentun transport) are independent of the magnetic Reynolds
number \citep[see also][]{F10}.

The remainder of the paper is organised as follows: in
Sect.~\ref{sec:model} we describe our model, and in
Sect.~\ref{sec:results} and \ref{sec:conclusions}, we present our
results and conclusions.

\section{The model}
\label{sec:model}
In an effort to keep the system as simple as possible, we assume that
the fluid is non-stratified and isothermal. The diffusion processes
are modeled with explicit Laplacian diffusion operators with constant
coefficients. A similar model was used by \cite{LKKBL09} and
\cite{KKV10}, although in these models higher order hyperdiffusive
operators were used instead of the Laplacian ones. The computational
domain is a cube with volume $H^3=(2\pi)^3$. We solve the usual set of
hydromagnetic equations in this geometry
\begin{eqnarray}
\frac{\mathcal{D} \bm{A}}{\mathcal{D} t} &=& -SA_y\hat{\bm{x}} - (\bm\nabla\bm{U})^{\rm T}\bm{A} - \eta \mu_0 \bm{J}, \\
\frac{\mathcal{D} \ln \rho}{\mathcal{D}t} &=& -\bm\nabla\cdot\bm{U}, \\
\frac{\mathcal{D} \bm{U}}{\mathcal{D}t} &=& -SU_x\hat{\bm{y}} -\cst^2{\bm \nabla}\ln\rho - 2\,\bm{\Omega} \times \bm{U} \nonumber \\ && \hspace{2cm}+ \frac{1}{\rho}(\bm{J} \times {\bm B} + \bm{\nabla} \cdot 2 \nu \rho \mbox{\boldmath ${\sf S}$}),\label{equ:mom}
\end{eqnarray}
where $\mathcal{D}/\mathcal{D}t=\pd/\pd t +
(\bm{U}+\meanv{U}^{(0)})\cdot\bm\nabla$ is the advective time
derivative, $\bm{A}$ is the magnetic vector potential, $\bm{B} =
\bm{\nabla} \times \bm{A}$ is the magnetic field, and $\bm{J} =
\mu_0^{-1} \bm{\nabla} \times \bm{B}$ is the current density, $\mu_0$
is the vacuum permeability, $\eta$ and $\nu$ are the magnetic
diffusivity and kinematic viscosity, respectively, $\rho$ is the
density, $\bm{U}$ is the velocity, and $\bm{\Omega}=\Omega_0(0,0,1)$
is the rotation vector. The large-scale shear is given by
$\meanv{U}^{(0)}=(0,Sx,0)$, with $q=-S/\Omega_0=1.5$, corresponding to
Keplerian rotation, in all runs. We use isothermal equation of state
$p=\cst^2 \rho$, characterised by a constant speed of sound,
$\cst$. In the present models we choose the sound speed so that the
Mach number remains of the order of 0.1 or smaller in order to
minimize the effects of compressibility. The rate of strain tensor
$\mbox{\boldmath ${\sf S}$}$ is given by
\begin{equation}
{\sf S}_{ij} = \onehalf (U_{i,j}+U_{j,i}) - \onethird \delta_{ij} \bm\nabla\cdot\bm{U},
\end{equation}
where the commas denote spatial derivatives.  The initial magnetic
field can be written in terms of the vector potential as
\begin{equation}
\bm{A}=A_0 \cos (k_{\rm A} x) \cos (k_{\rm A} z) \hat{\bm{e}}_y,
\end{equation}
where the amplitude of the resulting magnetic field that contains $x$
and $z$--components is given by $B_0=k_{\rm A}A_0$. We use $k_{\rm
  A}/k_1=1$, $\Omega_0={2\over3}\cdot10^{-1}\cst k_1$, and
$A_0={1\over3}\cdot10^{-1}\sqrt{\mu_0 \rho_0}\cst k_1^{-1}$ in all
models.

The values of $k_{\rm A}$, $\Omega_0$ and $A_0$ are selected so that
both the wavenumber with the largest growth rate, $k_{\rm
  max}=\Omega_0/u_{\rm A}=2$, where $u_{\rm A}=B_0/\sqrt{\mu_0
  \rho_0}$ is the Alfv$\acute{\rm e}$n velocity, and the largest
unstable wavenumber, $k_{\rm crit}=\sqrt{2q}k_{\rm max}\approx3.5$,
are well resolved by the grid. The other condition for the onset of
MRI, namely $\beta > 1$, where $\beta=2\mu_0 p/B_0^2$ is the ratio of
thermal to magnetic pressure, is also satisfied as $\beta=1800$ for
the maximum values of the initial magnetic field.

\begin{figure}
\begin{center}
\includegraphics[width=\columnwidth]{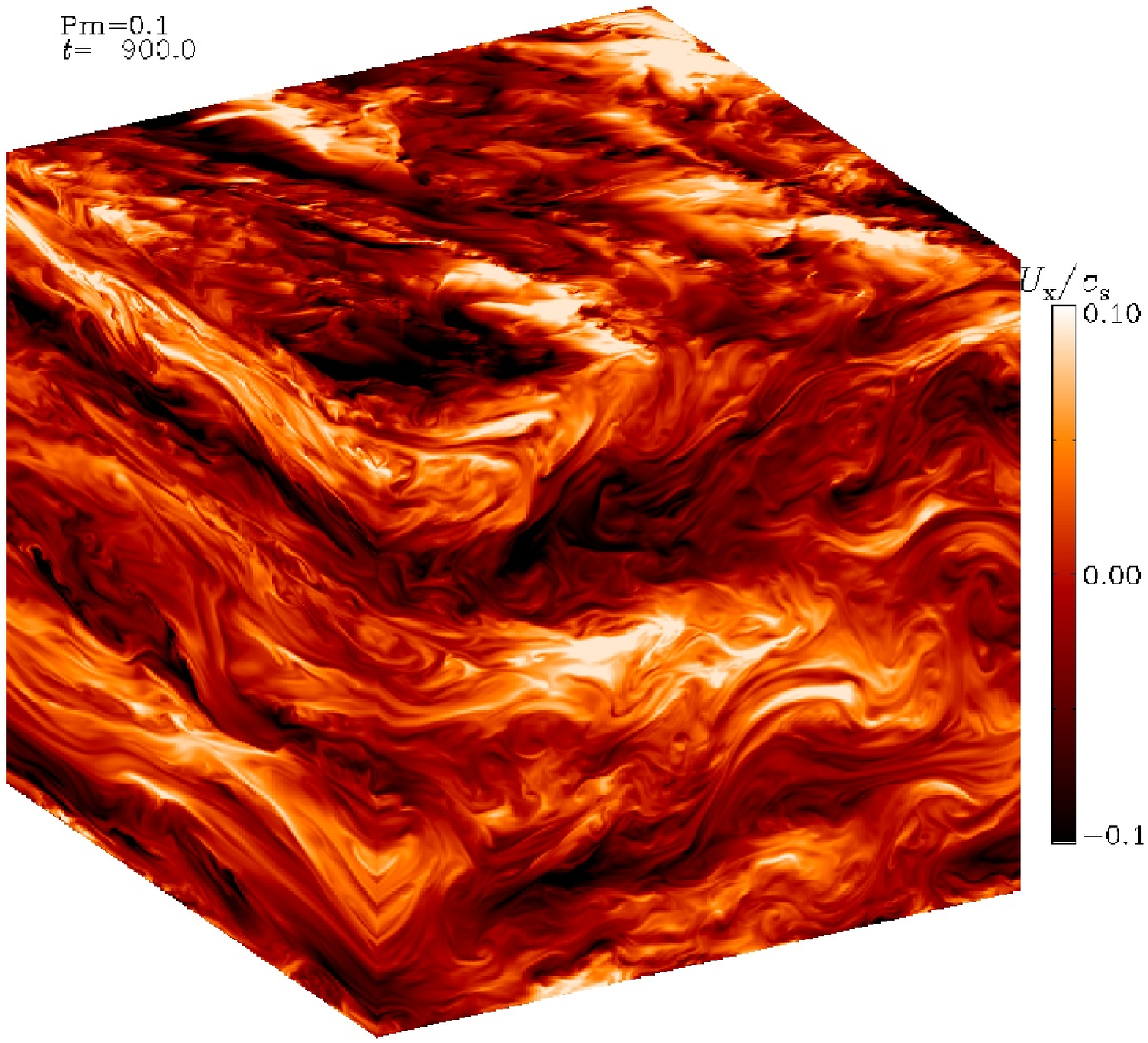}
\end{center}
\caption[]{Velocity component $U_x$ from the periphery of the domain for 
  Run~B9 with $\Pm=0.1$, $\Cem=1.5\cdot10^4$, and 
  $\Rey\approx9\cdot10^3$. See also
  \texttt{http://www.helsinki.fi/\ensuremath{\sim}kapyla/movies.html}
  for animations.
  }
\label{fig:vf512a1_Ux}
\end{figure}

We use the {\sc Pencil
  Code}\footnote{http://pencil-code.googlecode.com} which is a
high-order explicit finite difference method for solving the equations
of compressible magnetohydrodynamics. Resolutions of up to $512^3$
are used, see Figure~\ref{fig:vf512a1_Ux} for a snapshot of a high
resolution run.

\subsection{Boundary conditions}

In all models the $y$-direction is periodic and shearing-periodic
boundary conditions are used for the $x$-direction \citep{WT88}. On
the $z$-boundaries we use two sets of conditions. Firstly, we apply
periodic boundaries (denoted as PER).

Secondly, we apply a vertical field (VF) condition for the magnetic
field, which is fulfilled when
\begin{eqnarray}
B_x = B_y= B_{z,z}=0,
\end{eqnarray}
at the $z$-boundaries. 
In this case we use impenetrable,
stress-free conditions for the velocity according to
\begin{eqnarray}
U_{x,z} = U_{y,z} = U_z=0.
\end{eqnarray}
The novel property of the VF conditions is that they allow a net
toroidal flux to develop and allow magnetic helicity fluxes out of the
domain.

\begin{table*}
\caption{Summary of the runs. The Mach number (${\rm Ma}$) is 
  given by \Eq{equ:Mach},
  $\tilde\brms=\brms/B_{\rm eq}$, and 
  $\tilde{\mean{B}}_i=\sqrt{\mean{B}_i^2}/B_{\rm eq}$, 
  where $B_{\rm eq}$ is defined via \Eq{Beq}. 
  $\tilde{R}_{xy}=R_{xy}/(\Omega_0 H)^2$ and
  $\tilde{M}_{xy}=(\rho_0 \mu_0)^{-1}M_{xy}/(\Omega_0 H)^2$, 
  where $\rxy$ and $\mxy$ are 
  computed from equations~(\ref{equ:rxy}) and (\ref{equ:mxy}), respectively. 
  Finally, $\aSS$ is given by \Eq{equ:aSS}.}
\vspace{12pt}
\centerline{\begin{tabular}{lcccccccccccc}
    Run & grid  & $\Cem$ & $\Rem$ & $\Pm$ & $\rm Ma$ & $\tilde\brms$ & $\tilde{\mean{B}}_x$  & $\tilde{\mean{B}}_y$ & $\tilde{R}_{xy} [10^{-3}]$ & $\tilde{M}_{xy} [10^{-3}]$ & $\alpha_{\rm SS} [10^{-3}]$ & BC \\
    \hline
    A0  &  $128^3$ & $5\cdot10^3$   & --  & 5  &   --  &  --  & -- & -- & -- & -- & -- & PER    \\ 
    A1  &  $128^3$ & $10^4$         & 208 & 5  & 0.021 & 2.09 & 0.09 & 0.51 & 0.315 & $-2.162$ & $2.477\pm0.270$ & PER    \\ 
    A2  &  $128^3$ & $1.5\cdot10^4$ & 326 & 5  & 0.022 & 2.04 & 0.08 & 0.54 & 0.378 & $-2.337$ & $2.715\pm0.208$ & PER    \\ 
    A3  &  $256^3$ & $3\cdot10^4$   & 706 & 5  & 0.024 & 1.92 & 0.07 & 0.35 & 0.389 & $-2.564$ & $2.953\pm0.338$ & PER    \\ 
    \hline
    A4  &  $256^3$ & $3\cdot10^4$   & 377 & 2  & 0.013 & 1.78 & 0.04 & 0.31 & 0.102 & $-0.626$ & $0.728\pm0.212$ & PER    \\ 
    A5  &  $256^3$ & $6\cdot10^4$   & 625 & 2  & 0.010 & 1.83 & 0.04 & 0.33 & 0.079 & $-0.441$ & $0.520\pm0.074$ & PER    \\ 
    \hline
    A6  &  $256^3$ & $3\cdot10^4$   & 211 & 1  & 0.007 & 1.28 & 0.02 & 0.34 & 0.011 & $-0.075$ & $0.086\pm0.022$ & PER    \\ 
    A7  &  $256^3$ & $6\cdot10^4$   & 348 & 1  & 0.006 & 1.57 & 0.02 & 0.31 & 0.015 & $-0.088$ & $0.103\pm0.014$ & PER    \\ 
    \hline
    B0  &  $128^3$ & $1.5\cdot10^4$ &  -- & 20   &   --  &  --  &  --  & -- & -- & -- & -- & VF    \\ %
    B1  &  $128^3$ & $1.5\cdot10^4$ & 557 & 10   & 0.037 & 2.76 & 0.12 & 2.30 & 0.866 & $-4.726$ & $5.592\pm0.325$ & VF    \\ 
    B2  &  $128^3$ & $1.5\cdot10^4$ & 530 &  5   & 0.035 & 2.06 & 0.12 & 1.18 & 0.899 & $-4.802$ & $5.702\pm0.299$ & VF    \\ 
    B3  &  $128^3$ & $1.5\cdot10^4$ & 632 &  2   & 0.042 & 2.33 & 0.12 & 1.91 & 1.140 & $-4.577$ & $5.717\pm0.071$ & VF    \\ 
    B4  &  $128^3$ & $6.0\cdot10^3$ & 307 &  1   & 0.051 & 1.95 & 0.13 & 1.54 & 1.519 & $-5.463$ & $6.982\pm0.909$ & VF    \\ 
    B5  &  $128^3$ & $1.5\cdot10^4$ & 637 &  1   & 0.042 & 2.24 & 0.12 & 1.82 & 1.164 & $-4.422$ & $5.586\pm0.526$ & VF    \\ 
    B6  &  $256^3$ & $3.0\cdot10^4$ & 1242 & 1   & 0.041 & 1.77 & 0.11 & 0.97 & 1.018 & $-5.094$ & $6.111\pm0.560$ & VF    \\ 
    B7  &  $256^3$ & $1.5\cdot10^4$ & 687 & 0.5  & 0.046 & 1.69 & 0.12 & 1.04 & 1.154 & $-4.988$ & $6.142\pm0.636$ & VF    \\ 
    B8  &  $512^3$ & $1.5\cdot10^4$ & 719 & 0.2  & 0.048 & 1.55 & 0.11 & 0.87 & 1.111 & $-5.076$ & $6.187\pm1.068$ & VF    \\ 
    B9  &  $512^3$ & $1.5\cdot10^4$ & 897 & 0.1  & 0.060 & 1.78 & 0.12 & 1.39 & 1.680 & $-6.148$ & $7.828\pm1.335$ & VF    \\ 
    \hline
\label{Runs}\end{tabular}}\end{table*}

\subsection{Units, nondimensional quantities, and parameters}
Dimensionless quantities are obtained by setting
\begin{eqnarray}
k_1 = \cst = \rho_0 = \mu_0 = 1\;,
\end{eqnarray}
where $\rho_0$ is the mean density. The units of length, time,
velocity, density, and magnetic field are then
\begin{eqnarray}
&& [x] = k_1^{-1}\;,\;\; [t] = (\cst k_1)^{-1}\;,\;\; [U]=\cst\;,\;\; \nonumber \\ && [\rho]=\rho_0\;,\;\; [B]=\sqrt{\mu_0\rho_0\cst^2}\;. 
\end{eqnarray}
The simulations are controlled by the following dimensionless
parameters: the magnetic diffusion in comparison to viscosity is
measured by the magnetic Prandtl number
\begin{eqnarray}
\Pm=\frac{\nu}{\eta}.
\end{eqnarray}
The effects of viscosity and magnetic diffusion are quantified
respectively by the parameters
\begin{eqnarray}
\Cem=\frac{\cst}{\eta k_1^2}, \quad \frac{\Cem}{\Pm}=\frac{\cst}{\nu k_1^2}.
\end{eqnarray}
We also define the fluid and magnetic Reynolds numbers
\begin{eqnarray}
\Rey=\frac{\urms}{\nu k_1}, \quad \Rem=\frac{\urms}{\eta k_1}=\Pm\,\Rey,
\end{eqnarray}
where $\urms$ is the root-mean-square (rms) value of the velocity,
better decribing the nonlinear outcome of the simulations.
Furthermore, we often measure the magnetic field in terms of the
equipartition field which is defined via
\begin{eqnarray}
\Beq= \sqrt{\mu_0\langle \rho \urms^2}\rangle,\label{Beq}
\end{eqnarray}
where the brackets denote volume averaging. A convenient measure of
the turbulent velocity is the Mach number
\begin{eqnarray}
{\rm Ma}=\frac{\urms}{\cst}.\label{equ:Mach}
\end{eqnarray}
We define the mean quantites as horizontal averages
\begin{eqnarray}
\mean{F}_i(z,t)=\frac{1}{L_x L_y}\int_{-L_x/2}^{L_x/2} \int_{-L_y/2}^{L_y/2} F_i(x,y,z,t) dx dy.
\end{eqnarray}
Often an additional time average over the statically saturated state
is also taken. The size of error bars is estimated by dividing the
time series into three equally long parts.  The largest deviation of
the average for each of the three parts from that over the full time
series is taken to represent the error.

\begin{figure}
\begin{center}
\includegraphics[width=\columnwidth]{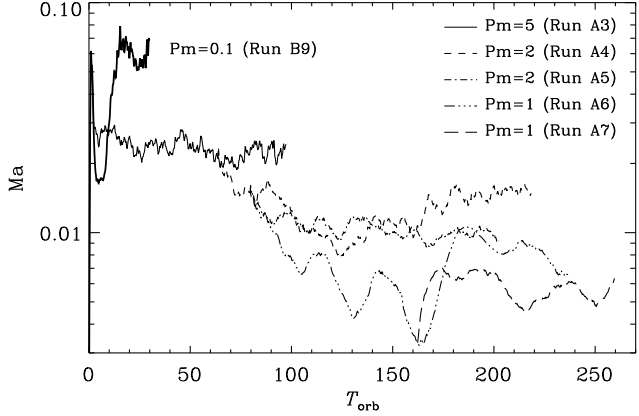}
\end{center}
\caption[]{Mach number defined via \Eq{equ:Mach} for
  Runs~A3--A7. The thick solid line shows the Mach number for Run~B9
  with $\Pm=0.1$ and VF boundaries.}
\label{fig:purmsA}
\end{figure}

\begin{figure}
\begin{center}
\includegraphics[width=\columnwidth]{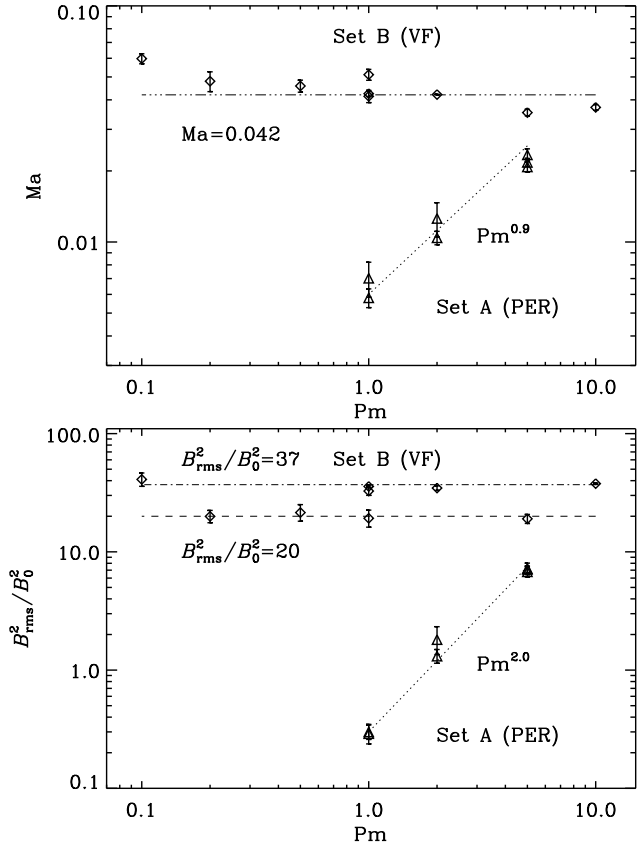}
\end{center}
\caption[]{Mach number (upper panel) and magnetic energy (lower panel)
  as functions of magnetic Prandtl number for periodic (triangles) and
  vertical field (diamonds) boundary conditions. The magnetic field is
  normalised by the rms value of the initial field.}
\label{fig:pubrms}
\end{figure}

\section{Results}
\label{sec:results}
We perform two sets of simulations listed in Table~\ref{Runs} where we
use either periodic (Set~A) or vertical field (Set~B) boundary
conditions. In Set~A, Runs~A0--A3 were started with the initial
conditions described in Section~\ref{sec:model} whereas Runs~A4--A6
were continued from a snapshot of Run~A3 in the saturated state, see
\Fig{fig:purmsA}. Run~A7 was continued from a snapshot of Run~A6 with
a two times lower diffusivities at roughly $150T_{\rm orb}$ where
$T_{\rm orb}=2\pi/\Omega_0$ is the orbital period. The minimum
duration of the runs in Set~A is $100T_{\rm orb}$. Runs in Set~B were
all started from scratch and typically ran a significantly shorter
time than those in Set~A, e.g.\ $\sim30T_{\rm orb}$ in the low--$\Pm$
cases (see \Fig{fig:purmsA}), because final saturation occurs much
faster.

\subsection{Saturation level of the MRI}

\subsubsection{Periodic case}
Earlier studies have shown that exciting the MRI in a periodic zero
net flux system becomes increasingly harder as the magnetic Prandtl
number is decreased \citep{FPLH07}. Furthermore, the saturation level
of turbulence has been reported to decrease as a function of
$\Pm$. This has been conjectured to be associated with the
difficulties of exciting a small-scale or fluctuation dynamo at low
$\Pm$ \citep[e.g.][]{Schekea07}. It is, however, unclear how the
saturation level of the small-scale dynamo is affected by this. It is
conceivable that at magnetic Reynolds numbers close to marginal it
takes a long time to reach saturation and that the current simulations
have not been run long enough. On the other hand, if catastrophic
quenching is to blame, the \emph{mean} magnetic field should decrease
as $\Rem^{-1}$ \citep[e.g.][and references therein]{BS05}. A further
possibility is the scenario suggested by \cite{V09}: in the absence of
an outer scale for the magnetic field, the microscopic diffusivities
determine the minimum lenght scale of MRI, which leads to turbulence
intensity decreasing proportional to $\Rem^{-2/3}$.

We study this issue by performing runs keeping $\Pm$ fixed and
increasing the Reynolds numbers. We find that the saturation level of
turbulence, measured by the Mach number and root mean square value of
magnetic field, are unaffected when $\Cem$ is increased by a factor of
three for the case $\Pm=5$ (Runs~A1--A3) and by a factor of two for
the cases $\Pm=2$ (Runs~A4--A5) and $\Pm=1$ (Runs~A6--A7), see
Table~\ref{Runs} and \Fig{fig:pubrms}. Furthermore, the Mach number
and rms magnetic field, normalised with the rms value of the initial
field, increase roughly linearly with $\Pm$. The $\Pm$-dependence of
rms magnetic field normalised to the equipartition field strength,
listed in Table~\ref{Runs}, shows a much weaker trend. This is to be
expected as $B_{\rm eq}$ is proportional to the rms velocity which, on
the other hand, is a produced by the magnetic field itself.  Since the
parameter range of our simulations is rather limited, no definite
conclusions can be drawn. However, taking the results at face value,
it appears that $\Pm$, not $\Cem$, is the parameter that determines
the saturation level in the periodic zero net flux case.  Recently,
\cite{F10} reached the same conclusion independently for the case of
$\Pm=4$.  According to our results, the catastrophic quenching and the
diffusivity-limited MRI length scale scenarios would be ruled
out. Although there is the possibility that our calculations have not
been run long enough, the results seem to suggest the small-scale
dynamo being harder to excite as $\Pm$ decreases.

\subsubsection{Vertical field case}

We find that the saturation behaviour is markedly different when
vertical field boundary conditions are applied (Table~\ref{Runs} and
\Fig{fig:pubrms}). The saturation level of turbulence depends only
weakly on the magnetic Prandtl number: the difference of the values of
$\urms$ between $\Pm=0.1$ and $\Pm=10$ cases is roughly 50 per cent.
Furthermore, the Mach number decreases as function of $\Pm$, the trend
being weaker but opposite to the periodic case.  This is likely caused
by the increase of viscosity by two orders of magnitude rather than
the intrinsic dependence of the MRI on $\Pm$. This conjecture is
supported by the saturation values of the magnetic fields which are
independent of $\Pm$ (lower panel of \Fig{fig:pubrms}). The runs in
Set~B, however, seem to fall into two distinct regimes of magnetic
field strength, where the magnetic energy differs by roughly a factor
of two. The reason for this apparent discrepancy is that a different
mode of the large-scale magnetic field is excited in the different
branches (see below). Similar behaviour of the large-scale dynamo has
previously been seen in isotropically forced turbulece \citep{BD02}.

\begin{figure}
\begin{center}
\includegraphics[width=\columnwidth]{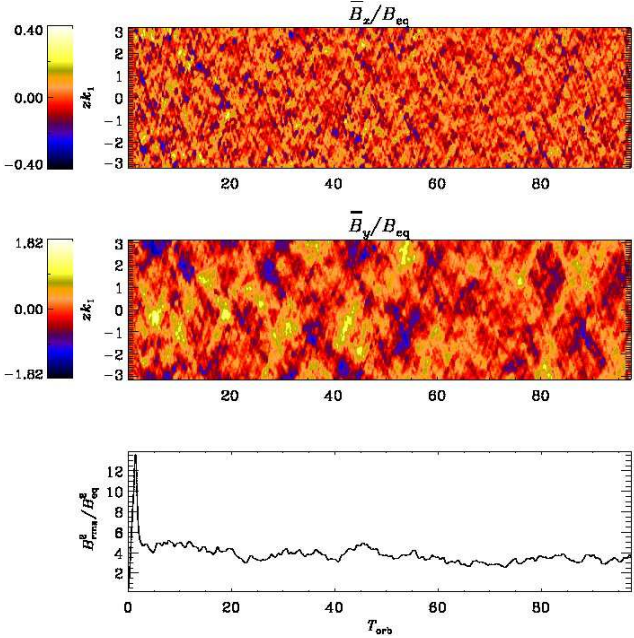}
\end{center}
\caption[]{Horizontally averaged horizontal magnetic fields
  $\mean{B}_x$ (top panel) and $\mean{B}_y$ (middle) for Run~A3 with
  $\Cem=3\cdot10^4$ and $\Pm=5$. The lower panel shows the square of
  the rms-value of the total magnetic field.}
\label{st_p256a}
\end{figure}

\begin{figure}
\begin{center}
\includegraphics[width=\columnwidth]{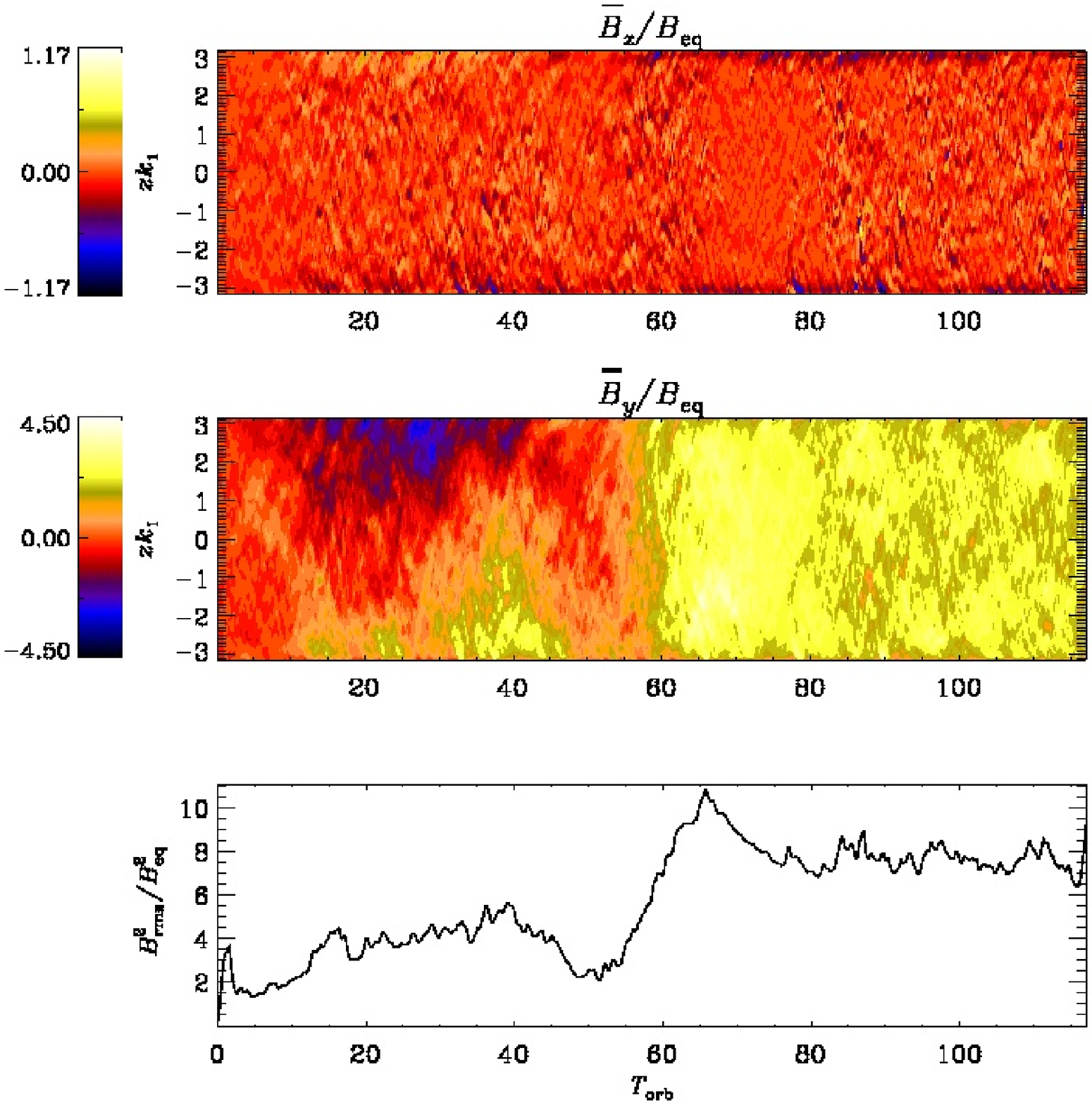}
\end{center}
\caption[]{Same as \Fig{st_p256a} but for Run~B1 with
  $\Cem=1.5\cdot10^4$ and $\Pm=10$.}
\label{st_vf128b3}
\end{figure}

\subsection{Large-scale magnetic fields}

In the runs with periodic boundaries we occasionally see the emergence
of large-scale magnetic fields with a sinusoidal dependence on $z$
(see \Fig{st_p256a}), i.e.\ $k/k_1=1$, in accordance with earlier
investigations \citep{LO08}. Similar large-scale dynamos have recently
been reported from nonhelically forced turbulence with shear where the
MRI is absent \citep[e.g.][]{Yetal08,BRRK08}. As in the forced
turbulence case a strong large-scale field is not present at all times
and the fields undergo apparently random sign changes that are not
fully understood \citep[see, however,][]{LO08,BRRK08}. The
intermittent nature of the large-scale fields could also explain the
apparent lack of catastrophical quenching of the time averaged mean
magnetic field (see Table~\ref{Runs}).

In the vertical field runs a strong large-scale dynamo is always
excited continuously. The two branches of solutions that are visible
in the total magnetic energy (\Fig{fig:pubrms}) are due to different
modes of the large-scale field. This is illustrated in
\Fig{st_vf128b3} where the horizontally averaged horizontal magnetic
field components are shown as functions of time for Run~B1. As is
common for dynamos with strong shear, the streamwise component of the
magnetic field is much stronger than the cross-stream one. Although
the initial condition of the magnetic field is the same in all runs,
the large-scale field which develops in the non-linear stage can
choose any of the available wavenumbers consistent with the vertical
boundary condition $B_x=B_y=0$. In practice, the dominant large-scale
component is $k/k_1=1$ or $k/k_1=\onehalf$ in our simulations. The
large-scale dynamo tends to accumulate energy at the smallest possible
wavenumber \citep{B01}, i.e.\ the largest spatial scale. However, if
the dominant mode is on some intermediate scale initially, those modes
can also be long-lived \citep{BD02}. Ultimately the large-scale field
evolves towards final saturation where the largest possible scale
dominates which was seen in \cite{BD02} and in some of our runs (cf.\
\Fig{st_vf128b3}). The fact that the magnetic energy in Runs~B2, B6,
B7, and B8 is smaller is due to the fact that the large-scale field is
predominantly of the $k/k_1=1$ flavour, and that final saturation of
the large-scale magnetic field has not yet occured. \cite{LO08} found
that the toroidal large-scale magnetic field generated in their
simulations is close to that yielding the maximum growth rate for an
$m=1$ non-axisymmetric instability. Using their notation we find a
similar result so that $\mean{B}_y k_y/(-S\sqrt{5/12} \sqrt{\mu_0
  \rho})\approx0.6$ for $k/k_1=1$ and $1.2$ for $k/k_1=\onehalf$,
using $k_y/k_1=1$ for the $m=1$ mode. However, the full signifigance
of this result is as of yet unclear.

Although the source of the turbulence and the nature of the dynamos
(kinematic vs.\ nonlinear) is different between the nonhelically
forced turbulence simulations \citep[e.g.][]{Yetal08,BRRK08} and the
non-stratified MRI runs such as those presented here, it is
conceivable that the large-scale field generation mechanism is the
same. Since the periodic system is homogeneous, the cause of the
large-scale fields cannot be the $\alpha$-effect of mean-field dynamo
theory \citep{M78,KR80}, which is in simple systems proportional to
the density gradient or the turbulence inhomogeneity due to boundaries
\citep[e.g.][]{GZR05,KKB10a}. However, a fluctuating $\alpha$ with
zero mean can also drive a large-scale dynamo when shear is present
\citep[e.g.][]{VB97,S97,S00,P07}.  This is the most likely source of
the large-scale magnetic fields in the present case. Furthermore, it
is possible that the shear--current and $\bm\Omega \times
\bm{J}$--effects can drive a large-scale dynamo \citep{R69,RK03,RK04},
although present evidence from numerical models does not support this
\citep{BRRK08}.

\begin{figure}
\begin{center}
\includegraphics[width=\columnwidth]{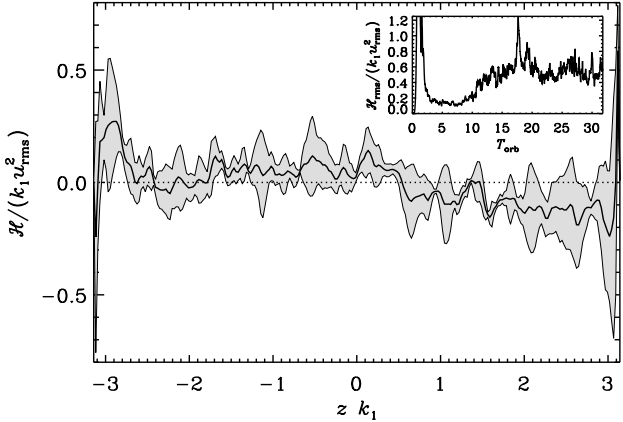}
\end{center}
\caption[]{Horizontally averaged kinetic helicity $\mathcal{H}$ from
  Run~B7. The inset shows the volume averaged rms-value of
  $\mathcal{H}$. The shaded area denotes the error estimates.}
\label{fig:pheli}
\end{figure}

In the VF runs the impenetrable stress-free $z$-boundaries make the
turbulence inhomogeneous near the boundary. This leads to the
generation of mean kinetic helicity
$\mathcal{H}(z)=\mean{\bm\omega\cdot\bm{u}}$, where
$\bm\omega=\bm\nabla\times\bm{u}$ is the vorticity.  The quantity
$\mathcal{H}$ is important, because the mean-field $\alpha$-effect is,
in simple settings, proportional to it \citep[e.g.][]{KR80}. Such
contributions, however, will not show up in volume averages because
the sign of the helicity, and thus of the $\alpha$-effect, are
different near the different boundaries. Figure~\ref{fig:pheli} shows
the horizontally averaged kinetic helicity for Run~B7. Here we average
also in time over the saturated state of the run. In most of the
volume the kinetic helicity is consistent with zero, although there
are regions close to the boundaries where non-zero mean values are
present. The rms-value of $\mathcal{H}$, however, is at least five
times greater than its mean (see the inset of \Fig{fig:pheli}). Note
also that the normalization factor contains the integral scale
$k_1$. A more proper definition would be to use the wavenumber where
turbulent energy peaks which is likely at least a factor of few
greater than $k_1$. Thus our estimates for the normalised helicity can
be considered as upper limits. The rather small values of mean
helicity and the dominance of fluctuations suggest that the generation
mechanism of the large-scale fields could indeed be the incoherent
$\alpha$--shear dynamo. However, a conclusive answer can only be
obtained by extracting the turbulent transport coefficients and by
performing mean-field modeling of the same system \citep[see
e.g.][]{G10}.

\begin{figure}
\begin{center}
\includegraphics[width=\columnwidth]{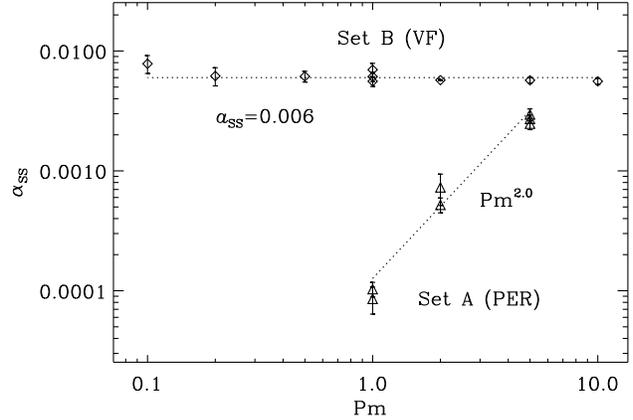}
\end{center}
\caption[]{Viscosity parameter $\aSS$ as a function of $\Pm$ for the
  runs listed in Table~\ref{Runs}. The dotted lines show $\aSS={\rm
    const}=6\cdot10^{-3}$ and $\aSS\propto\Pm^{2.0}$ for reference.}
\label{fig:palpha}
\end{figure}

\subsection{Angular momentum transport}

The main effect of turbulence in astrophysical disks is to enhance
diffusion which enables efficient accretion. In accretion disk theory
it is customary to parametrise the turbulent viscosity $\nut$ in
terms of the Shakura--Sunyaev viscosity parameter $\aSS$, which
relates $\nut$ with the local gas pressure \citep{SS73}.

We define the Shakura--Sunyaev viscosity parameter as \citep{BDH04}
\begin{eqnarray}
\alpha_{\rm SS}=\frac{[R_{xy}-M_{xy}/(\mu_0 \rho)]}{(\Omega_0 H)^2},\label{equ:aSS}
\end{eqnarray}
where
\begin{eqnarray}
R_{xy}\equiv \langle u_x u_y \rangle = \langle U_x U_y \rangle - \langle \mean{U}_x \mean{U}_y \rangle, \label{equ:rxy}
\end{eqnarray}
is the Reynolds stress and
\begin{eqnarray}
M_{xy}\equiv \langle b_x b_y \rangle = \langle B_x B_y \rangle - \langle \mean{B}_x \mean{B}_y \rangle, \label{equ:mxy}
\end{eqnarray}
the Maxwell stress, and where the angular brackets denote volume averaging. 
Here we decompose the velocity and
magnetic field into their mean ($\meanv{U}, \meanv{B}$),
taken here as the horizontal average, and fluctuating ($\bm{u},
\bm{b}$) parts.
The mean velocities show no systematic large-scale pattern and the
remaining signal $\mean{\bm{U}}\sim\mathcal{O}(0.05\urms)$ is likely a
residual of averaging over a finite number of cells. The contribution
of mean flows to the angular momentum transport and the dynamo process
is thus likely to be negligible.

For the runs in Set~A we find essentially the same scaling, consistent
with $\Pm^{2.0}$,
with magnetic Prandtl number as in the case of the turbulent kinetic
and magnetic energies, see \Fig{fig:palpha}. This is consistent
with the mixing length estimate of turbulent viscosity which is
proportional to the turbulence intensity \citep[e.g.][]{SKKL09}. The
numerical values of $\aSS$ decrease from $\approx10^{-3}$ for $\Pm=5$, to
$\aSS\approx 10^{-4}$ for $\Pm=1$.
In Set~B, on the other hand, $\aSS$ is essentially independent of
magnetic Prandtl number. The value of $\aSS$
is consistently of the order of $6\cdot10^{-3}$, which is significantly greater
than that found in runs with periodic boundaries. Here the qualitative
behaviour of $\aSS$ resembles that of the turbulent kinetic energy,
whereas the two different dynamo modes seen in magnetic energy are not
visible in the angular momentum transport.

\subsection{Discussion}
A possible clue to understanding the convergence problem in zero net
flux simulations comes from MRI models with density stratification: in
them the level of turbulence does converge when the Reynolds numbers
are increased \citep{DSP10}, even with perfect conductor of periodic
boundaries. Furthermore, such setups exhibit a large-scale dynamo
\citep[e.g.][]{BNST95,Sea96,G10} where the magnetic helicity changes
sign at the midplane \citep{G10}.

Recent numerical results from a different setting suggest that a
diffusive flux of magnetic helicity also exists \citep{MCCTB10}. Such
a flux can alleviate catastrophic quenching by transporting oppositely
signed magnetic helicity to the midplane where annihilation
occurs. This could explain the successful convergence of the
stratified MRI runs. In the non-stratified case with periodic or
perfectly conducting boundaries, however, no net flux of magnetic
helicity occurs and the large-scale dynamo can be catastrophically
quenched, shutting off the MRI. When a flux is allowed by changing to
vertical field boundary conditions, this limitation is removed and the
large-scale dynamo can operate without hindrance.  However, this
hypothesis requires further study and more careful analysis of the
helicity fluxes that we postpone to a future publication.

\section{Conclusions}
\label{sec:conclusions}

We present three-dimensional numerical simulations of the
magnetorotational instability in an isothermal non-stratified setup
with zero net flux initially. Using fully periodic boundaries, that do
not allow the generation of a mean toroidal flux or magnetic helicity
fluxes out of the system, we encounter the convergence problem
\citep{FPLH07} of the MRI: turbulent kinetic and magnetic energies,
and the angular momentum transport increase approximately proportional
to the magnetic Prandtl number. Intermittent large-scale magnetic
fields are observed in the periodic runs. Increasing the Reynolds
numbers moderately at a given $\Pm$ does not appear to markedly change
the results in the saturated state.

When vertical field boundary conditions, allowing the generation of a
mean flux and a magnetic helicity flux, are used, the MRI is excited
at least in the range $0.1\le\Pm\le10$ for our standard value of
$\Cem=1.5\cdot10^4$. We find that the saturation level of the
turbulence and the angular momentum transport are only weakly
dependent on the Prandtl number and that strong large-scale fields are
generated in all cases. The Shakura--Sunyaev viscosity parameter has
consistently a value of $\aSS\approx6\cdot10^{-3}$ in the vertical
field case. Exploring even lower values of $\Pm$ is infeasible at the
moment due to prohibitive computational requirements but there are no
compelling arguments against a large-scale dynamo operating at low
$\Pm$ \citep{B09}. We conjecture that the operation of the MRI at low
$\Pm$ is due to the efficient large-scale dynamo in the system. It is
conceivable that the dynamo only works if magnetic helicity is allowed
to escape \citep[see also][]{V09} or annihilate at the disk midplane
due to an internal diffusive flux \citep{MCCTB10}.  However, measuring
the magnetic helicity fluxes in the presence of boundaries is
difficult due to the fact that they are in general gauge dependent
\citep[e.g.][]{BDS02,HB10}.

The current results highlight the close connection between dynamo
theory and the theory of magnetised accretion disks \citep[see
also][]{Bl10} and the importance of studying the results in a common
framework \citep[e.g.][]{G10}. Clearly, a more thorough study is
needed in order to substantiate the possible role of magnetic
helicity fluxes for the excitation and saturation of the MRI. We plan
to address these issues in future publications.

\section*{Acknowledgments}
The authors acknowledge Axel Brandenburg for his helpful comments on
the manuscript. The numerical simulations were performed with the
supercomputers hosted by CSC -- IT Center for Science in Espoo,
Finland, who are administered by the Finnish Ministry of
Education. Financial support from the Academy of Finland grant Nos.\
121431 (PJK) and 112020 (MJK), are acknowledged. The authors
acknowledge the hospitality of NORDITA during the program ``Solar and
Stellar Dynamos and Cycles''.

\end{document}